\documentclass[aps,prd,twocolumn,showpacs,amsmath,amssymb]{revtex4}
  \usepackage{graphicx}

\begin{document}

\title{Vacuum Fluctuations of Energy Density can lead to the observed Cosmological Constant }

  \author{T. Padmanabhan  }
  \affiliation{ Inter-University Centre for Astronomy and Astrophysics, Post Bag
  4, Ganeshkhind, Pune-411 007.  email: nabhan@iucaa.ernet.in 
  }

\begin{abstract}
The energy density associated with Planck length is $\rho_{uv}\propto L_P^{-4}$ while the energy density associated with the Hubble length is $\rho_{ir}\propto L_H^{-4}$ where $L_H=1/H$. The observed value of the dark energy density
is quite different from {\it either} of these and is close to the geometric mean of the two: $\rho_{vac}\simeq \sqrt{\rho_{uv}
\rho_{ir}}$. It is argued that classical gravity 
is actually a probe of the vacuum {\it fluctuations} of energy density,  rather than the energy density itself.  While
the  globally defined ground state, being  an eigenstate of Hamiltonian,  will not have any fluctuations,
the ground state energy in the finite region of space bounded by the cosmic horizon will exhibit fluctuations $\Delta
\rho_{\rm vac}(L_P, L_H)$. When used as a source of gravity, this $\Delta \rho$ should lead to a spacetime
with a horizon size $L_H$. This bootstrapping condition leads naturally to an effective dark energy density
$\Delta\rho\propto (L_{uv}L_H)^{-2}\propto H^2/G$ which is precisely the observed value.  The model 
requires, either (i) a stochastic fluctuations of vacuum energy which is correlated over about a Hubble time or
(ii) a semi-anthropic interpretation.
 The implications are discussed.
\end{abstract}

\maketitle

The conventional discussion of the relation between cosmological constant and vacuum energy density is based on
evaluating the zero point energy of quantum fields with an ultraviolet cutoff and using the result as a 
source of gravity.
Any reasonable cutoff will lead to a vacuum energy density $\rho_{\rm vac}$ which is unacceptably high \cite{cc}. 

This argument,
however, is too simplistic since the zero point energy --- obtained by summing over the
$(1/2)\hbar \omega_k$ --- has no observable consequence in any other phenomena and can be subtracted out by redefining the Hamiltonian. The observed non trivial features of the vacuum state of QED, for example, arise from the {\it fluctuations} (or modifications) of this vacuum energy rather than the vacuum energy itself. 
This was, in fact,  known fairly early in the history of cosmological constant problem and, in fact, is stressed by Zeldovich \cite{zeldo} who explicitly calculated one possible contribution to {\it fluctuations} after subtracting away the mean value.
This
suggests that we should consider   the fluctuations in the vacuum energy density in addressing the 
cosmological constant problem.

Similar viewpoint arises, more formally, when we study the question of \emph{detecting} the energy
density using gravitational field as a probe.
 Recall that an Unruh-DeWitt detector with a local coupling $L_I=M(\tau)\phi[x(\tau)]$ to the {\it field} $\phi$
actually responds to $\langle 0|\phi(x)\phi(y)|0\rangle$ rather than to the field itself \cite{probe}. Similarly, one can use the gravitational field as a natural ``detector" of energy momentum tensor $T_{ab}$ with the standard coupling $L=\kappa h_{ab}T^{ab}$. Such a model was analysed in detail in ref.~\cite{tptptmunu} and it was shown that the gravitational field responds to the two point function $\langle 0|T_{ab}(x)T_{cd}(y)|0\rangle $. In fact, it is essentially this fluctuations in the energy density which is computed in the inflationary models \cite{stdinfl} as the seed {\it source} for gravitational field, as stressed in
ref.~\cite{tplp}. All these suggest treating the energy fluctuations as the physical quantity ``detected" by gravity, when
one needs to incorporate quantum effects. 

This fact alone, however, would not have helped. The vacuum expectation values in question are all ultraviolet
divergent and will --- normally --- scale as a suitable power of the UV-cutoff, $L_{uv}$.  If the energy density behaves as
$\rho\propto L_{uv}^{-4}$,  the fluctuation $\Delta\rho$ will also scale as $L_{uv}^{-4}$ since that is the dominant scale in the problem; so the cosmological constant will still come out too large. Another difficulty is that, at least formally, the ground state is an energy eigenstate and it will have no  dispersion in the energy. However, as we shall describe, there is a nice manner in which De-Sitter
spacetime circumvents  these difficulties and leads to the correct value for the cosmological constant.

The key new ingredient arises from the fact that the properties of the vacuum state (or, for that matter, any quantum state in field theory) depends on the scale at which it is probed. It is not appropriate to ask questions without specifying this scale. (This, in some sense, has been the key lesson from renormalisation group.) If the spacetime has a cosmological horizon which blocks information, the natural scale is provided by the size of the horizon,  $L_H$, and we should use observables defined within the accessible region. 
The operator $H(<L_H)$, corresponding to the total energy  inside
a region bounded by a cosmological horizon, will exhibit fluctuations  $\Delta E$ since vacuum state is not an eigenstate of 
{\it this} operator. The corresponding  fluctuations in the energy density, $\Delta\rho\propto (\Delta E)/L_H^3=f(L_{uv},L_H)$ will now depend on both the ultraviolet cutoff  $L_{uv}$ as well as $L_H$. 
We now see an interesting possibility of boot strapping: 
When used as the source of gravity, this $\Delta\rho$ should lead
to a spacetime with the horizon size $L_H$, which, in turn, requires us to 
compute $\Delta \rho$ using  $L_H$ as the infrared cutoff scale. 
Note that it is the existence of a cosmological \emph{horizon} at $L_H$ which provides a clear justification
for using this length scale in computing the energy fluctuations. This leads to a value for $\Delta \rho$
different from $L_{uv}^{-4}$ since we now have two length scales in the problem.
 This bootstrapping will lead to a relation between $\Delta\rho, L_{uv}$ 
and $L_H$. We will show that this relation is $\Delta\rho\propto (L_{uv}L_H)^{-2}\propto H^2/G$ if we take the
ultraviolet cut off at Planck length. That is precisely what we need.

 Remarkably enough, this result is very easy to obtain. 
 To obtain
 $\Delta \rho_{\rm vac} \propto \Delta E/L_H^3$ which scales as $(L_P L_H)^{-2}$
 we need to have $(\Delta E)^2\propto L_P^{-4} L_H^2$; that is, the square of the energy fluctuations
 should scale as the surface area of the bounding surface which is provided by the  cosmic horizon.  
 A simple argument to show that such a scaling is likely to occur is the following. Let the total Hamiltonian
 $H$ of the system be written as $H= H_1 + H_2$ where $H_1$ is the Hamiltonian obtained by
 integrating the Hamiltonian density within a sphere of radius $R ( = L_H) $ and $H_2$ denoting
 the Hamiltonian of the outside region. (We are dividing the real space rather than separating scales in the
momentum space, as is done in standard RG analysis.)
 Let $E_1$ and $E_2$ be the \emph{expectation values} of $H_1$ and $H_2$ in the ground state $|0\rangle $
 which is taken to be an eigen state of $H$ with $H|0\rangle  = E|0\rangle $.
 Consider now the energy
 dispersions in the regions 1 and 2. Writing $(H_2 - E_2)^2$ as $[ (H-E) - (H_1 - E_1) ]^2$ and noticing
 that the expectation values of $(H-E)$ and $(H-E)^2$ vanish in any energy eigenstate, we get the result
 \begin{equation}
 (\Delta E_2)^2 \equiv \langle 0|(H_2 - E_2)^2|0\rangle  = \langle 0|(H_1 - E_1)^2|0\rangle  = (\Delta E_1)^2
 \end{equation}
 That is, the dispersions in the energy in the regions 1 and 2 are equal.
 Since the regions 1 and 2 only share the bounding surface, this relation suggests that either dispersion
 could be proportional to the area of the surface and will scale as $R^2$. The dependence on $L_P$ 
 is then fixed by dimensional considerations and we get $(\Delta E)^2\propto L_P^{-4} R^2$.

 While the above argument is suggestive, it leaves significant scope for improvement --- not in the least 
 because one is dealing with formally divergent quantities. Fortunately, it is possible to do this 
 calculation rigorously. In flat spacetime, a scalar field has the familiar mode expansion
 \begin{equation}
 \phi (x) = \int \frac{d^3k}{(2\pi)^3} q_{\bf k} (t) e^{i{\bf k \cdot x}} 
 \label{mode}
\end{equation}
in terms of the harmonic oscillator modes $q_{\bf k}(t)$. Expressing the Hamiltonian in the region 1
\begin{equation}
H_1 = \int_{|x| < L_H} d^3{\bf x} \frac{1}{2} \left(\dot \phi^2 + |(\nabla \phi)|^2\right)
\end{equation}
 in terms of $\dot q_{\bf k} $ and $q_{\bf k}$ one can evaluate the dispersion 
\begin{equation}
(\Delta E )^2=\langle 0|(H_1 - \langle H_1\rangle )^2|0\rangle 
\end{equation}
 in a straightforward manner. To obtain a finite result, one needs to use an ultra violet regulator which 
 we take to be the Planck length.  For $L_H \gg L_P$, the final result has the scaling 
 \begin{equation}
 (\Delta E )^2 = c_1 \frac{L_H^2}{L_P^4} 
 \label{deltae}
 \end{equation}
 where the constant $c_1$ depends on the manner in which ultra violet cutoff is imposed.
 Similar calculations have been done (with a completely different motivation, in the context of 
 entanglement entropy)
 by several people and it is known that the area scaling  found in Eq.~(\ref{deltae}), proportional to $
L_H^2$, is a generic feature \cite{area}.
For a simple exponential UV-cutoff, $c_1 = (1/30\pi^2)$ but we do not believe this can be computed
 reliably without knowing the full theory. The result should also scale with the total number of degrees of freedom
at sub-Planckian energies which is not reliably known. [It is amusing to note that, the standard Zeldovich-Harrison
scale invariant spectrum has the mass fluctuation $(\delta M/M)^2\propto k^3P(k)\propto k^4\propto R^{-4}$
leading to the same ``area" scaling $(\delta M)^2\propto R^2$.]
 
 We thus find that the fluctuations in the energy density of the vacuum in a sphere of radius $L_H$ 
 is given by 
 \begin{equation}
 \Delta \rho_{\rm vac}  = \frac{\Delta E}{L_H^3} \propto L_P^{-2}L_H^{-2} \propto \frac{H^2}{G}
 \label{final}
 \end{equation}
 The numerical coefficient will depend on $c_1$ as well as the precise nature of infrared cutoff 
 radius (like whether it is $L_H$ or $L_H/2\pi$ etc.). It would be pretentious to cook up the factors
 to obtain the observed value for dark energy density.

 The current result in which the {\it fluctuations} in the energy density act as a source of gravity is
 similar in spirit to the idea presented  earlier in ref.\cite{tpcc}.  If a quantum  system is in a stationary state,
 its phase will evolve as $\exp(-iEt)$. But for a partial system which can undergo fluctuations
 in the energy, one cannot attribute such a phase  and we will --- instead ---
 obtain an uncertainty relation of the kind $\Delta E \propto (1/\tau)$ where $\tau$ is the 
 time scale of the fluctuations. Similarly, the semiclassical gravity will have a wave function with a phase 
$\exp (-i \rho_{vac} {\cal V})$ in case of a constant vacuum energy density, where ${\cal V}$ is
the four volume. In case of a finite region of space, the fluctuations lead to the result  
$\Delta \rho_{\rm vac} \propto (L_H L_P)^{-2}$.
 To be rigorous, we need to do this computation in the De-Sitter geometry rather than in flat spacetime.
 There is no conceptual difficulty in carrying out this program which merely requires replacing the mode functions
 in Eq.~(\ref{mode}) by the ones appropriate for De-Sitter spacetime. It is easy to see from dimensional 
 considerations that the results in  Eq.~(\ref{deltae}) and  Eq.~(\ref{final}) will continue to hold.
 The numerical factor changes and needs to be determined by numerical integration. 
 
We have concentrated on the energy density rather than on the full energy momentum tensor for the 
 sake of simplicity. More formally, one can construct the fluctuations in $T_{ab}$ along the same lines
 thereby obtaining the equation of state for the dark energy. If the regularization of the divergent expressions
 is handled in a Lorentz invariant manner (like, for example, using dimensional regularization in curved
 spacetime), then the Lorentz invariance dictates that the final result should have the form $\Delta T_{ab}=
 \Delta \rho_{\rm vac} g_{ab}$. 
 
The current observations (\cite{snmap}; for earlier indications of nonzero cosmological constant, see
\cite{early} ) do suggest that 
\begin{equation}
\rho_{vac}\simeq\sqrt{\rho_{UV}\rho_{IR}}
\label{key}
\end{equation}
if we take the UV scale as Planck scale and the IR scale as Hubble scale. This remarkable relation deserves an explanation which is provided in a simple manner by our analysis. In fact, if we start with this result and do a bit of 
``reverse-engineering", we immediately obtain the area scaling of energy fluctuations. The latter has been pursued 
 and obtained by several authors,  in connection with the holographic ideas and entropy-area connection.

The main criticism one could raise is regarding our assumption that we can ignore the mean energy density, regulated by the UV cut off to a large value and  concentrate on its fluctuation. 
In addition to the arguments given in the first paragraph of this paper,  there is another aspect which need to be stressed:  It is a fact of life that a fluctuation of magnitude $\Delta\rho_{vac}\simeq H^2/G$ will exist in the
energy density inside a sphere of radius $H^{-1}$ if Planck length is the UV cut off. {\it One cannot get away from it.}
On the other hand, observations suggest that there is a $\rho_{vac}$ of similar magnitude in the universe. It seems 
natural to identify the two, after subtracting out the mean value by hand. Our goal was more towards explaining why there is a surviving cosmological constant which satisfies Eq.(\ref{key}) which ---  in our opinion --- is {\it the} problem.  Another technical objection  could be regarding our computing the $\Delta E$ over a sphere of radius $H^{-1}$ and dividing it by the volume to get
$\Delta\rho_{vac}$. This corresponds to assuming the coherence scale of fluctuation to be Hubble radius, which is
indeed the correct thing to do. Recall that it is the horizon at $H^{-1}$ which motivates choosing this scale rather than some other smaller scale.

How does one interpret this result ? There are two possible ways. The first is to claim that the currently observed value of cosmological constant has no special significance and the fluctuations will lead to a $\rho_{vac}\approx H^2(t)/G$ at any epoch.
This suggests a stochastic interpretation of cosmological constant. To obtain the evolution of the universe, we now need to solve the Friedmann equations with $\rho_{vac}$ being a stochastic variable with 
a probability distribution
${\cal P}(\rho_{vac})$ which can also be computed in the ground state. Since the energy fluctuations
can be positive or negative, in general, the cosmological constant can have either sign. There are (at least!) two difficulties in this interpretation: 

(i) For this idea to be viable,
we need the fluctuations to be correlated over a time scale of the order of $H(t)^{-1}$. It is trivial to compute the
two point function $C(t)=\langle 0|H_1(t_1)H_1(t_1+t)|0\rangle$ for the quantum fluctuations and one finds that
$C(t)=c_2(L_H^2t^{-4})$; this quantity is finite and we do not need to introduce the UV cutoff. (If the cutoff is introduced
the result is $C(t)\propto  L_H^2(L_P^2+t^2)^{-2}$). Being a power law correlation, this suggests fluctuations at all time scales with the scale governed by $L_H$. But when $t\approx L_H$, the ratio
$[C(L_H)/C(0)]^{1/2}=(L_P/L_H)^2\ll 1$  {\it if $C(0)$ is computed with a cut off.}. This probably means that
coherence over $t\approx L_H$ is rare. But the argument is not conclusive since what is required is the coherence time, {\it given that a fluctuation of order $L_H^2/L_P^4$ has occurred}. If we assume that coherence over 
$t\approx L_H$ is possible, then
the current positive value of cosmological constant
is a result of a fluctuation lasting typically for a time scale of $H_0^{-1}$ and we would predict to see a reversal
of sign of $\rho_{vac}$ at earlier epochs. (Except that, it is difficult to predict whether this should have occurred at $z=2$ or at
$z=2\pi^2$, say, with any confidence until all the numerical factors are accounted for.
On the other hand, if the SN data at a redshift of 2, say, does indicate a negative cosmological constant, ones faith in this stochastic fluctuations will increase!). With this interpretation, the model solves both the conventional ``problems"
of the cosmological constant. Its value is correctly explained as due to energy fluctuations and this relation holds at any epoch over a local time scale of $H(t)^{-1}$ thereby answering ``why now?" by ``always!".

(ii) The second difficulty is is not special to our model and  arises whenever one tries to model quantum fluctuations as a classical stochastic process and treat it as a source of semiclassical gravity. In the standard cosmological model, the consistency of $(^0_0)$ component of Einstein's equation with the
$(^1_1)$ component will demand that $d(\rho a^3)=-p d(a^3)$ which in turn will require $\rho$ to be independent of time if $p=-\rho$. This relation will require a reinterpretation when cosmological constant is treated as a stochastic variable. If the coherence time of the fluctuation is of the order of Hubble time, there will be constancy of effective
$\rho_{vac}$ (and standard evolution of the universe) within one Hubble time. But when stochastic fluctuation of cosmological constant occurs one needs to interpret the energy conservation equation in some limiting form. 

Other issues that one could raise regarding the computation in this approach are relatively straight forward to address: one can work out the probability distribution for fluctuations in $T_{ab}$ in a De-Sitter background and integrate the stochastic Friedman equation. One also need to study the effect of the cosmological constant fluctuations on the growth of structures. All of these are best addressed numerically since one need to deal with a
stochastic process with finite correlation.

There is a completely different way of interpreting this result based on some imaginative ideas suggested by Bjorken
\cite{bjorken} recently. The key idea here is to parametrise the universes by the value of $L_H$ which they have. It is a fixed, pure number for each universe in an ensemble of universes but all the other parameters of the physics are correlated with
$L_H$. This is motivated by a series of arguments in ref. \cite{bjorken} and, in this approach, $\rho_{vac}\propto L_H^{-2}$ almost by definition; the hard work was in determining how other parameters scale with $L_H$. In the approach suggested here, a dynamical interpretation of the scaling  $\rho_{vac}\propto L_H^{-2}$ is given as due to vacuum fluctuations of other fields. We now reinterpret each member of of the ensemble of universes as having zero energy density for vacuum (as any decent vacuum should have) but the effective $\rho_{vac}$ arises from the quantum fluctuations {\it with the correct scaling}. One can then invoke standard anthropic-like arguments (but with very significant differences as stressed in ref. \cite{bjorken} ) to choose a range for the size of our universe. This appears to be much more attractive way of interpreting the result than introducing stochastic, time dependent fluctuations.

Finally, to be fair, this attempt should be
 judged in the backdrop of other suggested solutions almost all of which require introducing
 extra degrees of freedom in the form of scalar fields, modifying gravity or introducing higher dimensions etc. {\it and} fine tuning the potentials (for a non-comprehensive sample of 
 references, see \cite{sample}.) At a fundamental level such approaches are unlikely to provide the final solution.

I thank K.Subramanian for several rounds of extensive discussions and R.Nityananda and A.D.Patel for useful 
comments on previous drafts.

\end{document}